\documentclass[aps,prb,twocolumn,showpacs,groupedaddress]{revtex4}
\usepackage{graphics}

\usepackage{dcolumn}
\begin{document}

\title
{Superconducting transition in disordered granular superconductor in magnetic 
fields
}

\author{Ryusuke Ikeda}

\affiliation{%
Department of Physics, Kyoto University, Kyoto 606-8502, Japan
}

\date{\today}

%%%%%%%% &&& soft spin version no details wo dousuruka ? 
%%% 3d de large fluct. low T de large disorder 
%%% sono riyuu: \xi_0^2 ga 3d de 0.385 bai 2d ni kurabe, smaller
%%% keisuu 4d no sei. 
%%% granular deha strong quantum fluctuation de Hvg << Hc2 at high T demo, 
%%% strong disorder. Soft spin version 
%%%  

\begin{abstract}
Motivated by a recent argument that the superconducting (SC) transition field of three-dimensional (3D) disordered superconductors with granular structure in a nonzero magnetic field should lie above $H_{c2}(0)$ in low $T$ limit, the glass transition (or, in 2D, crossover) curve $H_g(T)$ of disordered quantum Josephson junction arrays is examined by incorporating SC fluctuations. It is found that the glass transition or crossover in the granular materials can be described on the same footing as the vortex-glass (VG) transition in amorphous-like (i.e., nongranular) materials. 
In most of 3D granular systems, the vanishing of resistivity upon cooling should occur even above $H_{c2}(0)$, while the corresponding sharp drop of the resistivity in 2D case may appear only below $H_{c2}$ as a result of an enhanced quantum fluctuation. 
\end{abstract}


%\keyword{granular superconductor, vortex-glass, quantum phase transition}

\maketitle

\section{Introduction}

Throughout extensive studies on phase diagrams of nongranular systems \cite{NS,Bla,FFH}, it is understood at present that the superconducting transition, characterized by the vanishing of resistivity, in homogeneously disordered (amorphous-like) type II superconductors under nonzero fields occurs as a glass transition. As far as the static disorder is point-like, the resulting glass transition curve $H_g(T)$ at nonzero temperatures ($T > 0$) does not deviate much from the melting transition line of a clean vortex lattice. For amorphous-like ({\it nongranular}) three-dimensional (3D) systems, the $H_g(T)$ curve, determined resistively, approaches a field near $H_{c2}(0)$ \cite{com1} in low $T$ limit, even if it {\it apparently} approaches a field range below $H_{c2}(0)$ \cite{Okuma} upon cooling as the reduced temperatures $T/T_{c0}$ is not low enough. On the other hand, It is believed that the glass transition in 2D disordered case occurs only \cite{FFH,MPF} at $H_{\rm SI} = H_g(T=0)$ below $H_{c2}(0)$ and corresponds to a field-tuned superconductor-insulator (SI) 
transition at $T=0$ \cite{MPF,IR,RI1}. 

However, it is unclear whether this picture also holds in {\it granular} systems or not. By neglecting effects of vortex pinning and superconducting (SC) fluctuations, a glass phase {\it peculiar to} disordered granular superconductors was obtained \cite{JL} as a phase lying {\it above} $H_{c2}(T)$ defined \cite{com0} at longer scales than the intergrain spacing (see Fig.1). Hereafter, this glass phase, which may appear even in $H=0$ separately from the Meissner state, will be called as the phase glass (PG). However, the fate of PG is not clear once SC fluctuation and vortex pinning effects are taken into account to describe {\it real} systems. It was argued recently within a standard model with dissipative quantum phase dynamics that $H_g(T)$ of 3D disordered {\it granular} systems near $T=0$ should lie far above $H_{c2}(0)$ \cite{GaL,Gesh} and that, even in 2D granular systems with no genuine glass transition \cite{FFH} in $T > 0$, the corresponding crossover line $H_g(T)$ defined from a sharp drop of the resistance should show, upon cooling, a divergent (upward) low $T$ behavior as if it terminates at a field $H_g(0)$ higher than $H_{c2}(0)$ \cite{GaL} in low $T$ limit (see Fig.2). This argument may be consistent with the presence of PG in the mean field phase diagram Fig.1 if the PG is superconducting. However, if so, it is unclear how the SC (glass) phase and the portion of $H_g(T)$ in $H < H_{c2}$ in Fig.2 are described. Further, the argument in Ref.\cite{GaL} for 2D case is incompatible with the field-tuned S-I transition behavior which is believed to occur below $H_{c2}(0)$ even in granular SC thin films \cite{Gant}. 
%%%%%%%%%%%%%%%%%%%
\begin{figure}[t]
\scalebox{2.0}[2.0]{\includegraphics{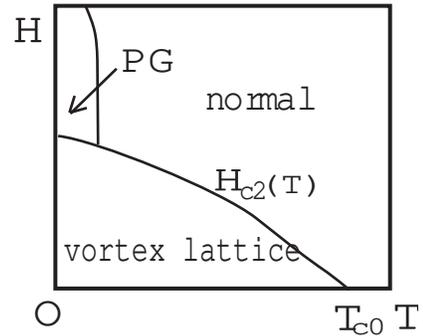}}
\caption{{\it Mean-field} \cite{JL} $H$-$T$ phase diagram of a disordered granular superconductor.} \label{fig.1}
\end{figure}
%%%%%%%%%%%%%%%%%
%%%%%%%%%%%%%%%%%%%
\begin{figure}[t]
\scalebox{2.0}[2.0]{\includegraphics{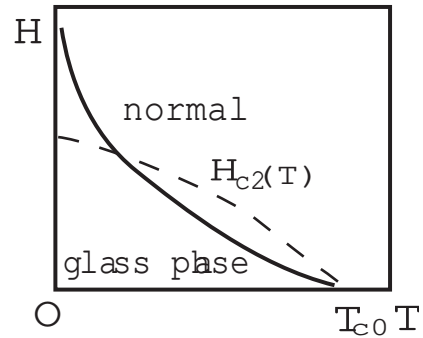}}
\caption{$H$-$T$ phase diagram, proposed phenomenologically \cite{GaL}, of a disordered granular superconductor. In 3D case, the resistance vanishes on $H_g(T)$ (solid curve). } \label{fig.2}
\end{figure}
%%%%%%%%%%%%%%%%%

In this paper, the glass transition curve $H_g(T)$ of disordered granular superconductors is examined in the mean field approximation but by including SC fluctuations. We show that $H_g(T)$ in the granular case is obtained in formally the same manner as the vortex-glass (VG) transition curve in the amorphous-like case and hence, is the SC transition curve in 3D systems at which the resisitivity vanishes \cite{FFH,RI2} at least in type II limit. 
%at the mean field level \cite{com4}. 
Further, we find that, even if the mean field PG phase is absent even at $T=0$, a phase diagram of the type of Fig.2 with $H_g(0) > H_{c2}(0)$ is generically obtained in 3D, while the situation in which $H_g(0) < H_{c2}(0)$ is easily reached in 2D, reflecting an enhanced quantum fluctuation. We argue that the main origin of $H_g(0)$ lying above $H_{c2}(0)$ is an {\it enhancement} of vortex pinning due to the SC fluctuation. 
%Due to the latter result, if the PG phase be a metal, it would have become diff%icult to find a proper phase diagram near $T=0$ (see Fig.1 (b)). 

In sec.II, a model of disordered and dissipative Josephson-junction array is introduced and rewritten into an effective action, and $H_g(T)$ curves in 2D and 3D cases are derived in sec.III. For comparison, the VG transition curve in 3D amorphous case is given in the same framework as in IV, and related discussions and a summary are given in sec.V. 

\section{Model}
We start from the hamiltonian 
\begin{equation}\label{eq:1} 
{\cal H}_\theta = \alpha \sum_j \biggl(-{\rm i} \frac{\partial}{\partial {\hat \theta}_j} \biggr)^2 - \sum_{<i, j>} J_{ij} \, {\rm cos}({\hat \theta}_i - {\hat \theta}_j), 
\end{equation}
describing a Josephson junction array with a charging energy $2 \alpha$ on each grain, where the pair of indices, $<i$, $j>$, denotes a nerest-neighbor pair of sites (i.e., grains), and ${\hat \theta}_j$ is a phase operator on the $j$-th grain. Below, the model will be extended to a more general one including effects of possible dissipation on each grain and of electromagnetic fields. The most straightforward method of performing this is to express the model (\ref{eq:1}) into the corresponding quantum action 
\begin{eqnarray}\label{eq:2}
{\cal S}\! &=& \! {\cal S}_0 \!- \! \int^\beta_0 \! d\tau \! \sum_{<i,j>} \biggl[ \frac{J_{ij}}{2} \exp[{\rm i}(\theta_{i}(\tau) - \theta_{j}(\tau) \\ \nonumber &-& e^* A_{{\rm ex}, i-j} - e^* \delta A_{i-j}(\tau))] 
+ c.c. \biggr] \\ \nonumber 
&+& \int d\tau \int d\tau' \sum_j \frac{\nu}{\pi (\tau - \tau')^2} [ \, 1 - {\rm cos}(\theta_j(\tau) - \theta_j(\tau')) \, ] 
\end{eqnarray}
in the unit $\hbar=c=1$, where $\beta=1/T$, $e^*$ is the Cooper-pair charge, $A_{{\rm ext},i-j}$ denotes the line-integral of an external gauge field over the bond $i-j$, $\delta A_{i-j}(\tau)$ is the corresponding gauge disturbance introduced for obtaining the conductivity in the $i-j$ direction, and 
\begin{eqnarray}\label{eq:3}
{\cal S}_0(\theta) &=& \int^\beta_0 d\tau \sum_i \frac{1}{4 \alpha} \biggl(\frac{\partial \theta_i(\tau)}{\partial \tau} \biggr)^2
\end{eqnarray}
is the action corresponding to the charging energy, i.e., the first term of eq.(1). 

Note that the dissipative (last) term of eq.(\ref{eq:2}) is expressed as 
\begin{equation}\label{eq:4}
{\cal S}_{\rm dis} = \beta^{-1} \sum_j \sum_\omega \frac{\nu}{2} |\omega| 
|\Phi_j(\omega)|^2,
\end{equation}
where $\Phi_j(\omega)$ is the Fourier transform of 
\begin{equation}\label{eq:5}
\Phi_j(\tau) = \exp({\rm i} \theta_j(\tau)). 
\end{equation}
That is, eq.(\ref{eq:4}) is nothing but the familiar dissipative term, written in the phase-only approximation, in the time-dependent Ginzburg-Landau model. 

The quenched disorder in the system is incorporated into a randomness of $J_{ij} = J_{ji}^*$ with a nonzero real mean $J_0$, i.e., ${\overline {J_{ij}}}=J_0 > 0$,  and a Gaussian distribution ${\overline {(J_{ij} - J_0) (J_{ji} - J_0)}} = J^2$. These relations may be regarded as being due to a {\it random} gauge field $a_{ij}$ defined by $J_{ij}-J_0 \propto \exp({\rm i} a_{ij})$. 
The free energy $F = - \beta^{-1} {\rm ln} Z$ will be expressed in terms of the replica trick as $F = - \beta^{-1} (Z^n - 1)/n$ in $n \to +0$ limit. The averaged replicated partition function ${\overline {Z^n}}$ is given by \cite{Illcon} 
\begin{equation}\label{eq:6}
{\overline {Z^n}} = {Z_0^n} \, < \exp(- {\cal S}_f - {\cal S}_g) 
>_0, 
\end{equation}
where $Z_0$ is the partition function of ${\cal S}_0$, $< \, \, \, >_0$ denotes the ensemble average on $\sum_{1 \leq a \leq n} {\cal S}_0(\theta^{(a)})$, and 
\begin{eqnarray}\label{eq:7}
{\cal S}_f &=& - \sum_{a=1}^n \sum_{<i,j>} \int_0^\beta \! d\tau J_0 \, {\rm cos}( e^* \delta A_{i-j}(\tau) + e^* A_{{\rm ex},i-j} \\ \nonumber 
&-& \theta_i^{(a)}(\tau) + \theta_j^{(a)}(\tau) ) 
+ \int d\tau_1 \int d\tau_2 \sum_j \frac{\nu}{(\tau_1 - \tau_2)^2} 
\\ \nonumber 
&\times& [ \, 1 - {\rm cos}(\theta_j^{(a)}(\tau_1) - \theta_j^{(a)}(\tau_2)) \, ], \\ \nonumber 
{\cal S}_g &=& - \frac{1}{2} \int d\tau \int d\tau' \sum_{a,b} \sum_{<i,j>}  J^2 \cos(e^* (\delta A_{i-j}(\tau) \\ \nonumber 
&-& \delta A_{i-j}(\tau')) + \theta_i^{(a)}(\tau) - \theta_i^{(b)}(\tau') - \theta_j^{(a)}(\tau) + \theta_j^{(b)}(\tau')). 
\end{eqnarray}
Before proceeding further, ${\cal S}_f$ will be rewritten in the form \cite{Otterlo} 
\begin{eqnarray}\label{eq:8}
{\cal S}_f &=& {\rm const}. - d J_0 \beta^{-1} \sum_\omega \sum_{i, a} (\Phi^{(a)}_i(\omega))^* \biggl( 1 - \frac{\nu}{2 d J_0} |\omega| \! \! \! \! \\ \nonumber 
&+& \frac{1}{2d} {\bf D}_i \cdot {\bf D}^*_i \biggr) \Phi_i^{(a)}(\omega) 
\\ \nonumber 
&\simeq& {\rm const}. - d J_0 \beta^{-1} \sum_\omega \, [ 1 + \nu |\omega|/(2 d J_0) ]^{-1} \sum_{i,a} (\Phi^{(a)}_i(\omega))^* \nonumber \\
&\times& \biggl( 1 + \frac{1}{2d} {\bf D}_i \cdot {\bf D}^*_i \biggr) \, \Phi^{(a)}_i(\omega) \\ \nonumber
\end{eqnarray}
for the cubic or square lattice in $d$-dimension, where $\Phi_i(\tau) = \beta^{-1} \sum_\omega \Phi_i(\omega) e^{-{\rm i}\omega \tau}$. Equation (9) is valid up to the lowest order in $ \nu |\omega|/J_0$ and the laplacian ${\bf D}_i \cdot {\bf D}^*_i/(2d)$, and ${\bf D}_i$ is the gauge-invariant gradient on the lattice \cite{Kleinert} accompanied by the gauge field ${\bf A}_{\rm ex} + \delta {\bf A}(\tau)$. 
Then, by introducing the conventional SC order parameter $\psi^{(a)}_i(\tau)$ and the glass order parameter $q^{(ab)}_i(\tau_1,\tau_2)=(q^{(ba)}(\tau_2, \tau_1))^*$, ${\overline Z^n}$ becomes \cite{JL,Illcon,Read} 
\begin{equation}\label{eq:9}
\frac{\overline {Z^n}}{\overline {Z^n_0}} = \int {\cal D}\psi^{(a)} {\cal D}(\psi^{(a)})^* {\cal D} q^{(ab)} \exp(-{\cal S}_{\rm eff}(\psi, q)), 
\end{equation}
where 
\begin{eqnarray}\label{eq:10}
&{\cal S}_{\rm eff}&(\psi, q) = \int d\tau_1 \! \! \int d\tau_2 \sum_{a,b} \sum_{i} \frac{J^{-2}}{2} q^{(ab)}_i(\tau_1, \tau_2) \\ \nonumber 
&\times& q^{(ba)}_i(\tau_2, \tau_1) 
+ \frac{\beta^{-1}}{4d} \sum_\omega \sum_i \sum_a (1 + \frac{\nu}{2d J_0} |\omega|) |\psi_{i, \omega}^{(a)}|^2 \\ \nonumber
&-& \sum_i {\rm ln} \biggl[ \biggl< T_\tau \exp \biggl(\frac{\sqrt{J_0}}{2} \int d\tau \sum_a \Phi_{i}^{(a)}(\tau) \biggl( 1 \\ \nonumber 
&+& \frac{{\bf D}_{i} \cdot {\bf D}_{i}^*}{2d} \biggr)^{1/2} 
(\psi_i^{(a)}(\tau))^* 
+ \frac{1}{2} \int d\tau_1 d\tau_2 \sum_{a,b} \Phi_{i}^{(a)}(\tau_1) \\ \nonumber 
&\times& (\Phi_{i}^{(b)}(\tau_2))^* \biggl( 1 + \frac{{\tilde {\bf D}}_{i} \cdot {\tilde {\bf D}}_{i}^*}{2d} \biggr)^{1/2} q_i^{(ba)}(\tau_2, \tau_1) 
+ {\rm c.c.} \biggr) \biggr>_0 \biggr],  
\end{eqnarray}
where 
\begin{equation}\label{eq:14}
\psi(\tau) = \beta^{-1} \sum_\omega \psi_\omega e^{-{\rm i}\omega \tau}, 
\end{equation}
and ${\tilde {\bf D}}_i$ denotes the gauge-invariant gradient on the lattice accompanied by the gauge field $\delta {\bf A}(\tau_1) - \delta {\bf A}(\tau_2)$. 
Performing the cumulant expansion in powers of $q^{(ab)}$ and $\psi^{(a)}$ in the logarithmic term, various terms such as 
\begin{eqnarray}\label{eq:11}
\int d\tau_1 \int &d\tau_2& <T_\tau \Phi^{(a)}(\tau_1) (\Phi^{(a)}(\tau_2))^*>_0 \! q^{(aa)}(\tau_1,\tau_2), \\ \nonumber
\frac{1}{2} \int &d\tau_1& \int d\tau_2 \int d\tau_3 \int d\tau_4 <T_\tau \Phi^{(a)}(\tau_1) (\Phi^{(a)}(\tau_3))^* \\ \nonumber 
\times \Phi^{(b)}(&\tau_4&) (\Phi^{(b)}(\tau_2))^* >_0 q^{(ba)}(\tau_2,\tau_1) q^{(ab)}(\tau_3,\tau_4),
\end{eqnarray} 
arise in the resulting Landau action ${\cal S}_{\rm eff}$ appropriate to the ensuing analysis. The average $<\,\,\,>_0$ is carried out by using ${\cal S}_0$ or its soft-spin version \cite{com5}. For instance, $<T_\tau \Phi_i^{(a)}(\tau_1) (\Phi_i^{(a)}(\tau_2))^*>_0$ becomes $\exp(-\alpha|\tau_1 - \tau_2|/2)$ in the low $T$ limit. Below, it will be replaced by its local limit $4 \delta(\tau)/\alpha$ anywhere except in the lowest order term in $q^{(ab)}$. Further, the $T$-dependence will be taken into account just in the $|\psi|^2$ term because, at least, one of such $T$-dependences is necessary in order to keep a reasonable mean field $H_{c2}(T)$ line for the $\psi$-field. It will be clear that these simplifications are not essential to the present purpose of addressing the low $T$ phase diagram. 

Next, let us write $q^{(ab)}(\tau_1, \tau_2)$, by following Read et al. \cite{Read}, as $Q^{(ab)}(\tau_1, \tau_2) - C \delta_{a,b} \delta(\tau_1-\tau_2)$ in order to delete the term $\int d\tau_1 \int d\tau_2 |Q^{(ab)}(\tau_1, \tau_2)|^2$. By representing spatial coordinates in terms of the continuous coordinates ${\bf x}$, we finally obtain the following effective Landau action 
\begin{eqnarray}\label{eq:12}
t &{\cal S}_{\rm eff}&(\psi, Q; \delta {\bf A}) =  \int \frac{d^d{\bf x}}{a^d} \biggl[ \int \frac{d\tau}{\kappa} \sum_a \biggl( \frac{\partial^2}{\partial \tau_1 \, \partial \tau_2} + r \biggr) \\ \nonumber 
&\times& Q^{(aa)}({\bf x}; \tau_1, \tau_2) \biggr|_{\tau_1=\tau_2} - \frac{\kappa}{3} \int d\tau_1 d\tau_2 d\tau_3 \sum_{a,b,c} Q^{(ab)}({\bf x}; \tau_1,\tau_2) \\ \nonumber 
&\times& Q^{(bc)}({\bf x}; \tau_2,\tau_3) Q^{(ca)}({\bf x}; \tau_3, \tau_1) + \frac{u}{2} \int d\tau \sum_a (Q^{(aa)}({\bf x}; \tau,\tau))^2  \\ \nonumber
&+& \frac{t a^2}{4 d \alpha^2} \sum_{a,b} \int d\tau_1 \int d\tau_2 |(-{\rm i}\nabla - e^*(\delta {\bf A}(\tau_1) - \delta {\bf A}(\tau_2))) \\ \nonumber 
&\times& Q^{(ab)}({\bf x}; \tau_1, \tau_2)|^2 \biggr] + t {\tilde {\cal S}}_{\rm eff}, 
\end{eqnarray}
where 
\begin{eqnarray}\label{eq:13}
t &{\tilde {\cal S}}_{\rm eff}& = a^{-d} \int d^d{\bf x} \biggl[ \sum_a \biggl( \beta^{-1} \sum_\omega (d_\psi |\omega| |\psi^{(a)}_\omega|^2 ) \\ \nonumber 
&+& \int d\tau \biggl[ r_{\psi, 0} |\psi^{(a)}(\tau)|^2 + c_\psi \biggl|\frac{\partial \psi^{(a)}}{\partial \tau} \biggr|^2 \\ \nonumber 
&+& t \, {\tilde a}^2 |(-{\rm i}\nabla - e^* {\bf A}_{\rm ex} - e^* \delta {\bf A}(\tau) ) \psi^{(a)}(\tau)|^2 \\ \nonumber 
&+& \frac{t}{2 \alpha} \biggl(\frac{u_R}{\alpha} \biggr) \biggl(\frac{4 J_0}{\alpha} \biggr)^2 |\psi^{(a)}({\bf x}, \tau)|^4 \biggr ] \biggr) \\ \nonumber
&-& w_\psi \sum_{a,b} \int d\tau_1 \int d\tau_2 (\psi^{(a)}({\bf x}, \tau_1))^* Q^{(ab)}({\bf x}; \tau_1, \tau_2) \\ \nonumber 
&\times& \psi^{(b)}({\bf x}, \tau_2) 
 \biggr].
\end{eqnarray}
Here, we have introduced a short length cut-off $a$ which corresponds to the intergrain spacing. We assume that $a$ is much longer than the coherence length of the host material forming the grains, and hence that the averaged $H_{c2}(0)$ of the granular system is lower than the {\it microscopic} $H_{c2}(0)$ of the host material forming {\it each} grain \cite{com0}. The $H_{c2}(0)$ mentioned in sec.I is nothing but this averaged $H_{c2}(0)$. This is consistent with the assumption in choosing the phase-only model (1) as a starting model that the amplitude of the pair-field in {\it each} grain be robust. 

We note that, although the dissipative term in eq.(\ref{eq:1}) is reflected only in the term quadratic in $\psi$ of the effective action, the dynamics of the glass fluctuation $\delta Q^{(ab)}$ also becomes dissipative through the coupling ($w_\psi$-) term between $\psi$ and $Q^{(ab)}$ after integrating over the SC ($\psi$-) fluctuations. 

Using the soft-spin version of the zero-dimensional action ${\cal S}_0$, i.e., a Ginzburg-Landau action corresponding to ${\cal S}_0$ \cite{com5}, we find the coefficients to be given by 
\begin{eqnarray}\label{eq:15}
t &=& \frac{\alpha^3}{4}, \\ \nonumber
\cr \kappa &=& 2, \\ \nonumber 
\cr r &=& \frac{\alpha^2}{4} \biggl( \biggl(\frac{\alpha}{2J} \biggr)^2 + 1 \biggr) \biggl( \biggl(\frac{\alpha}{2J} \biggr)^2 - 3 \biggr), 
\\ \nonumber 
\cr u &=& 4 \frac{\alpha \, u_R}{J^2} \biggl( 1 - \frac{2 J^2}{\alpha^2} \biggr), \\ \nonumber 
\cr \frac{r_{\psi,0}}{t} &=& \frac{1}{4d} - \frac{J_0}{2 \alpha} + \frac{J_0 T}{2 \alpha^2}- \frac{J_0}{4 \alpha} \biggl( \biggl(\frac{\alpha}{2J} \biggr)^2 - 1 \biggr), \\ \nonumber 
\cr \frac{c_\psi}{t} &=& \frac{J_0}{2 \alpha^3}, \\ \nonumber 
\cr d_\psi &=& \frac{t \nu}{8 d^2 J_0}, \\ \nonumber 
\cr {\tilde a}^2 &=& a^2 \frac{J_0}{4 d \alpha}, \\ \nonumber 
\cr w_\psi &=& \frac{J_0 t}{\alpha^2}. 
\end{eqnarray} 
The coefficient $u_R$ denotes the {\it renormalized} four-point vertex in the soft-spin version of ${\cal S}_0$. In low dimensional cases with $d < 2$, the renormalization of the fluctuation in a classical Ginzburg-Landau action is well approximated \cite{Scalapino} by the Hartree approximation in which $u_R=0$. Based on this fact, the vertex $u_R/\alpha$ of the quantum zero-dimensional action ${\cal S}_0$ may be assumed to take a (dimensionless) number much less than 
unity. 

Below, the Fourier transform of the glass field $Q^{(ab)}({\bf x})$ is defined, by following Ref.\cite{Read}, as 
\begin{eqnarray}\label{eq:16}
Q^{(ab)}(\tau_1,\tau_2; {\bf x}) 
&=& q^{(ab)} 
 + \frac{1}{\beta} \sum_{\omega \neq 0} {\overline D}_\omega e^{-{\rm i}\omega(\tau_1-\tau_2)} \, \delta_{a,b} \\ \nonumber 
&+& \beta^{-2} \sum_{\omega_1, \omega_2} \delta Q^{(ab)}_{\omega_1, \omega_2}({\bf x}) e^{-{\rm i}\omega_1 \tau_1 - {\rm i}\omega_2 \tau_2}, 
\end{eqnarray}
where the replica symmetric form 
\begin{equation}\label{eq:rs}
q^{(ab)} = q (1 - \delta_{a,b}) + {\overline q} \, \delta_{a,b}
\end{equation}
is assumed for $q^{ab}$, because we do not study here the glass phase below $H_g$. Further, we focus hereafter on the situation that $H_g$ is approached from the higher temperatures at which $q=0$, and ${\overline q}=\beta^{-1} {\overline D}_0$ \cite{Read}. The replica-diagonal component ${\overline D}_\omega$ is the Fourier transform of the (imaginary) time correlation between two "spins" (Re$\Phi_j$, Im$\Phi_j$) and hence, is nonvanishing even above $H_g(T)$. It is determined by the variational equation $0= n^{-1} \partial {\overline {Z^n}}/\partial 
{\overline D}_\omega$, or 
\begin{equation}\label{eq:Dw}
\kappa {\overline D}_\omega^2 = \kappa^{-1} (\omega^2 + r)  + u \beta^{-1} \sum_\omega {\overline D}_\omega - w_\psi \beta^{-1} \langle \psi^*_\omega({\bf x}) \psi_\omega({\bf x}) \rangle_s, 
\end{equation}
where $\langle \, \, \, \, \rangle_s$ denotes the ensemble {\it and} space averages. 
The physically meaningful solution of eq.(\ref{eq:Dw}) is 
\begin{equation}\label{eq:Dw1}
{\overline D}_\omega = - \kappa^{-1} \, \sqrt{ {\tilde r}_\omega + \omega^2},
\end{equation}
where 
\begin{eqnarray}\label{eq:Dw2}
{\tilde r}_\omega \!= r + u \, \kappa \, \beta^{-1} \sum_\omega {\tilde D}_\omega - w_\psi \, \kappa \, \beta^{-1} \langle \psi^*_\omega({\bf x}) \psi_\omega({\bf x}) \rangle_s. 
\end{eqnarray}
The minus sign of eq.(\ref{eq:Dw1}) is chosen so that a physically correct "spin" correlation along the (imaginary) time direction is recovered \cite{Read}. We note that the last term of eq.(\ref{eq:Dw2}) is nonvanishing in $T \to 0$ 
limit. 

\section{Glass Transition Line}

In this section, we will determine the glass transition field $H_g(T)$ of the granular system described by eqs.(\ref{eq:12}) and (\ref{eq:13}). It will be seen below that, upon cooling, the glass transition is described as a vortex-glass ordering \cite{FFH,RI2} induced by the coupling between the SC fluctuation $\psi$ and the glass fluctuation $\delta Q^{(ab)}$. 

We will focus on the (highest) instability temperature at which the glass fluctuation $\delta Q^{(ab)}$ becomes critical at the Gaussian level when the SC fluctuation is fully incorporated. It corresponds to a mean field glass transition in analogy to the normal to Meissner mean field transition following from the BCS theory (Note that the quasiparticle in the normal state in the latter corresponds to the SC fluctuation in the former). For simplicity, we shall identify this mean field glass transition line with $H_g(T)$. Further, since our main purpose here is to give a correct answer on $H_g(T)$ in high $H$ and low $T$ portion of the $H$-$T$ phase diagram, we will use the lowest Landau level (LLL) approximation for the $\psi$ modes. For the LLL modes, the operation $(-{\rm i}\nabla_\perp - e^* {\bf A}_{\rm ext})^2$ is replaced by $|e^*| H$, and the number of field-induced vortices may be expressed as the total magnetic flux $H S$ multiplied by $|e^*|/2 \pi$, where $\sqrt{S}$ is the linear system size. 

Under the nonzero ${\overline D}_\omega$ given by eq.(\ref{eq:Dw1}), the glass fluctuation $\delta Q^{(ab)}$ obeys the following effective action $\delta {\cal S}_{\rm eff} = \delta {\cal S}_Q 
+ \delta {\cal S}_\psi$ up to the quadratic order in $\delta Q^{(ab)}$, where 
\begin{eqnarray}\label{eq:26}
\delta {\cal S}_Q &=& \frac{\beta^{-2}}{2} \sum_{a,b} \sum_{\omega_1, \omega_2} a^{-d} \int d^d{\bf x} \biggl[ J^{-2} a^2 \nabla \delta Q^{(ab)}(\omega_1, \omega_2) \\ \nonumber 
&\times& \cdot \nabla \delta Q^{(ba)}(-\omega_2, - \omega_1) \\ \nonumber 
&-& \frac{\kappa}{t} ({\overline D}_{\omega_1} + {\overline D}_{\omega_2}) \delta Q^{(ab)}(\omega_1, \omega_2) \delta Q^{(ba)}(-\omega_2, - \omega_1) \\ \nonumber 
&-& 2 \frac{J_0}{\alpha^2} (\psi^{(a)}_{\omega_1})^* \delta Q^{(ab)}(\omega_1, -\omega_2) \psi^{(b)}_{\omega_2} \biggr], \\ \nonumber 
\delta {\cal S}_\psi &=& \beta^{-1} t^{-1} a^{-d} \int d^d{\bf x} \sum_a \biggl[ \sum_{\omega} \biggl[ (r_{\psi,0} - J_0 \alpha^{-2} t {\overline D}_\omega \\ \nonumber 
&+& c_\psi \omega^2 + d_\psi |\omega|)|\psi^{(a)}_\omega|^2 
+ \frac{J_0 t a^2}{4 d \alpha} |(-{\rm i}\nabla - e^*{\bf A}_{\rm ex}) \psi^{(a)}_\omega|^2 \biggr] \\ \nonumber 
&+& \sum_{\omega_i} \frac{t u_R J_0^2}{2 \beta^2} \biggl(\frac{2}{\alpha}\biggr)^4 \delta_{\omega_1+\omega_2, \omega_3+\omega_4} (\psi_{\omega_1}^{(a)} \psi_{\omega_2}^{(a)})^* \psi^{(a)}_{\omega_3} \psi^{(a)}_{\omega_4} \biggr].
\end{eqnarray}
The $u \, \delta Q^{(aa)} \delta Q^{(aa)}$ term (see eq.(\ref{eq:12})) was dropped from $\delta {\cal S}_Q$. In fact, this term acts \cite{Read} as an {\it interaction} term between $\delta Q$s and hence, may be neglected at the present stage of focusing on the noninteracting (Gaussian) $\delta Q$-fluctuation. In addition, since $u \, \alpha \ll \sqrt{r}$ for any $J/\alpha$ values of our interest, this term is quantitatively negligible even in obtaining ${\overline D}^{(d)}_\omega$ (see eq.(\ref{eq:Dw2})), where ${\overline D}^{(d)}_\omega$ denotes ${\overline D}_\omega$ in $d$-dimension. Hence, let us drop this small term consistently hereafter. Then, the $T$ and $H$ dependences in ${\overline D}^{(d)}_\omega$ arises primarily from the SC fluctuation. 

First, let us give the renormalized $\psi$-fluctuation in order to reasonably describe the field range lower than $H_{c2}(T)$. Hereafter, the $|\psi|^4$ term will be treated, as in the nongranular case \cite{Dorsey}, in the Hartree approximation. The Hartree approximation in LLL may be invoked as an infinite-range limit of a nonlocal Ginzburg-Landau model \cite{Hikami}. Although the use of the nonlocal model makes the details of vortex positional ordering obscure \cite{IOT}, the description of the glass ordering in strongly-disordered superconductors is not essentially affected by this procedure. Then, the $\psi$-propagator in LLL is defined in 2D ($d=2$) case by 
\begin{equation}
\beta^{-1} \langle \psi^*_\omega \psi_\omega \rangle_s = h^{(2)} G^{(2)}_{\rm dia}, 
\end{equation}
where 
\begin{equation}\label{eq:30}
h^{(2)} = \frac{|e^*| H a^2}{2 \pi}, 
\end{equation}
\begin{equation}\label{eq:27}
G_{\rm dia}^{(2)}(\omega) = \frac{\alpha}{J_0 \mu_\omega^{(2)}},  
\end{equation}
\begin{eqnarray}\label{eq:28}
\mu_\omega^{(2)} &=& \frac{\alpha}{J_0} \biggl( \frac{c_\psi}{t}\omega^2 + \frac{d_\psi}{t}|\omega| + \frac{r_{\psi, 0}}{t} \biggr) + \frac{2 \pi h^{(2)}}{8} \\ \nonumber &-& \alpha^{-1} {\overline D}_\omega^{(2)} + \Sigma^{(2)}, 
\end{eqnarray}
and 
\begin{equation}\label{eq:29}
{\overline D}_\omega^{(2)} = - \kappa^{-1} \biggl( r+\omega^2 - h^{(2)} \frac{\kappa t}{\alpha \mu_\omega^{(2)}} \biggr)^{1/2}.
\end{equation}
In 3D, the corresponding expressions are given by replacing $G^{(2)}_{\rm dia}$ with $a \int dk_z/(2 \pi) G^{(3)}_{\rm dia}(k_z)$, where 
\begin{equation}\label{eq:31}
G_{\rm dia}^{(3)}({k_z}; \omega) = \frac{\alpha}{J_0(\mu_\omega^{(3)} 
+ k_z^2 a^2 /12)}, 
\end{equation}
\begin{eqnarray}\label{eq:32}
\mu_\omega^{(3)} &=& \frac{\alpha}{J_0} \biggl( \frac{c_\psi}{t}\omega^2 + \frac{d_\psi}{t}|\omega| + \frac{r_{\psi, 0}}{t} \biggr) + \frac{2 \sqrt{3} \pi h^{(3)}}{36} \\ \nonumber &-& \alpha^{-1} {\overline D}_\omega^{(3)} 
+ \Sigma^{(3)},  
\end{eqnarray}
\begin{equation}\label{eq:33}
{\overline D}_\omega^{(3)} = - \kappa^{-1} \biggl( r+\omega^2 - h^{(3)} \frac{\kappa t}{\alpha (\mu_\omega^{(3)})^{1/2}} \biggr)^{1/2}, 
\end{equation}
and 
\begin{equation}\label{eq:34}
h^{(3)} = 3^{1/2} \frac{|e^*| H a^2}{2 \pi}. 
\end{equation}
The self energies $\Sigma^{(2)}$ and $\Sigma^{(3)}$ due to the interaction ($u_R$-) term are given by 
\begin{equation}\label{eq:35}
\Sigma^{(2)} = 32 \frac{u_R}{\alpha} (\beta \alpha)^{-1} h^{(2)} \sum_\omega 
(\mu_\omega^{(2)})^{-1},
\end{equation}
and 
\begin{equation}\label{eq:36}
\Sigma^{(3)} = 32 \frac{u_R}{\alpha} (\beta \alpha)^{-1} h^{(3)} \sum_\omega (\mu_\omega^{(3)})^{-1/2}. 
\end{equation}
It is easily verified that, in $J \ll \alpha$, eqs.(\ref{eq:28}) and (\ref{eq:32}) reduce to their results in the $J=0$ case. 

Although solving exactly these set of equations for each $d$ is in general difficult, it can be performed just at the (mean-field) glass transition line $H_g(T)$. To define $H_g(T)$, let us first rewrite eq.(\ref{eq:26}) into an effective action $\delta {\cal S}_{{\rm eff}, Q}$ consisting only of $\delta Q$. Within the Hartree approximation for the $\psi$-fluctuation, $\delta {\cal S}_{{\rm eff}, Q}$ in 3D becomes 
\begin{eqnarray}\label{eq:37}
\delta {\cal S}_{{\rm eff}, Q} &\simeq& \frac{1}{2} \int_{\bf k} \sum_{a,b} \sum_{\omega_1, \omega_2} | \delta Q^{(ab)}_{\bf k}(\omega_1,\omega_2) |^2 \! \! \! \! \! \\ \nonumber 
&\times& \biggl( J^{-2} k^2 - \frac{\kappa}{t} ({\overline D}_{\omega_1}^{(3)} + {\overline D}_{\omega_2}^{(3)}) \\ \nonumber 
- &h^{(2)}& \biggl(\frac{J_0}{\alpha^2} \biggr)^2 v_{{\bf k}_\perp} \int \!\frac{dq}{2 \pi} G_{\rm dia}^{(3)}(q; \omega_1) G_{\rm dia}^{(3)}(q + k_3; -\omega_2) \biggr), 
\end{eqnarray} 
where $v_{{\bf k}_\perp} = \exp(-(k_1^2+k_2^2)/(2 h^{(2)}))$. Then, by focusing on the term with $\omega_j=0$ and ${\bf k}=0$ in eq.(\ref{eq:37}), 
$H_g(T)$ is defined as 
\begin{equation}\label{eq:38}
-2 t^{-1} {\overline D}^{(3)}_{\omega=0} = \frac{h^{(3)}}{2 \kappa \alpha^2} (\mu^{(3)}_{0})^{-3/2}
\end{equation}
in 3D, while 
\begin{equation}\label{eq:39}
-2 t^{-1} {\overline D}^{(2)}_{\omega=0} = \frac{h^{(2)}}{\kappa \alpha^2} (\mu^{(2)}_{0})^{-2} 
\end{equation}
in 2D, where $\mu^{(d)}_0 \equiv \mu^{(d)}_{\omega=0}$. To rewrite these equations more explicitly, $\mu^{(d)}_\omega$ at $H_g(T)$ will be expressed as $\mu^{(d)}_{0} + \delta \mu^{(d)}_\omega$. If using eq.(\ref{eq:28}) or (\ref{eq:32}), it is not difficult to obtain $\delta \mu_\omega^{(d)}$ as a function of $\mu_{0}^{(d)}$, and, up to the lowest order in $|\omega|$, it becomes 
\begin{eqnarray}\label{eq:40}
\delta \mu_\omega^{(3)} = 2 \biggl( \frac{\mu_0^{(3)}}{3 + 64 (\mu_0^{(3)})^{5/2}/h^{(3)}} \, \frac{d_\psi \alpha}{t J_0} |\omega| \biggr)^{1/2} 
%&\delta \mu_\omega^{(3)}& = 2 \biggl( \frac{\mu_0^{(3)}}{3 + 64 (\mu_0^{(3)})^{%5/2}/h^{(3)}} \biggr)^{1/2} \\ \nonumber
%&\times& \biggl( \frac{d_\psi \alpha}{t J_0} |\omega| + \biggl( \frac{c_\psi \a%lpha}{t J_0} - \frac{1}{2 \kappa^2 \alpha {\overline D}_{\omega=0}^{(3)}} \bigg%r)\omega^2 \biggr)^{1/2} 
\end{eqnarray}
in 3D, and 
\begin{eqnarray}\label{eq:41}
\delta \mu_\omega^{(2)} = 2 \biggl( \frac{\mu_0^{(2)}}{1 + 8 (\mu_0^{(2)})^{3}/h^{(2)}} \, \frac{d_\psi \alpha}{t J_0} |\omega| \biggr)^{1/2}
%&\delta \mu_\omega^{(2)}& = 2 \biggl( \frac{\mu_0^{(2)}}{1 + 8 (\mu_0^{(2)})^{3%}/h^{(2)}} \biggr)^{1/2} \\ \nonumber 
%&\times& \biggl( \frac{d_\psi \alpha}{t J_0} |\omega| 
% + \biggl( \frac{c_\psi \alpha}{t J_0} - \frac{1}{2 \kappa^2 \alpha {\overline %D}_{\omega=0}^{(2)}} \biggr)\omega^2 \biggr)^{1/2} 
\end{eqnarray}
in 2D, respectively. 
The expressions on $\mu_\omega^{(d)}$ obtained above imply that, to the lowest order in the Matsubara frequency, the glass fluctuation propagator just at the (mean field) glass transition takes the form 
\begin{equation}\label{eq:42}
\langle \delta Q^{(ab)}_{\bf k}(\omega, \omega') \delta Q^{(ba)}_{-{\bf k}}(-\omega', -\omega) \rangle \simeq (k^2 + |\omega|^{1/2} + |\omega'|^{1/2})^{-1} 
\end{equation}
after being rescaled spatially. 
%Thus, the dynamical exponent of this mean field glass transition is four. 

Now, by applying the above expressions of $\delta \mu^{(d)}_\omega$ to the selfconsistent equations on $\mu_0^{(d)}$ 
\begin{eqnarray}\label{eq:43}
\mu_{0}^{(2)} = \frac{\alpha}{J_0 t} r_{\psi, 0} + \frac{2 \pi h^{(2)}}{8} - \alpha^{-1} {\overline D}_{\omega=0}^{(2)} + \Sigma^{(2)}, 
\end{eqnarray}
and 
\begin{eqnarray}\label{eq:44}
\mu_{0}^{(3)} = \frac{\alpha}{J_0 t} r_{\psi, 0} + \frac{2 \sqrt{3} \pi h^{(3)}}{36} - \alpha^{-1} {\overline D}_{\omega=0}^{(3)} + \Sigma^{(3)}, 
\end{eqnarray}
we obtain the coupled equations, eqs.(\ref{eq:28}), (\ref{eq:29}), (\ref{eq:35}), (\ref{eq:39}), (\ref{eq:41}), and (\ref{eq:43}) in 2D and eqs.(\ref{eq:32}), (\ref{eq:33}), (\ref{eq:36}), (\ref{eq:38}), (\ref{eq:40}), and (\ref{eq:44}) in 3D. The resulting $H$-$T$ relation for each $d$ is nothing but the $H_g(T)$ line. 

%%%%%%%%%%%%%%%%%%%
\begin{figure}[t]
\scalebox{0.35}[0.35]{\includegraphics{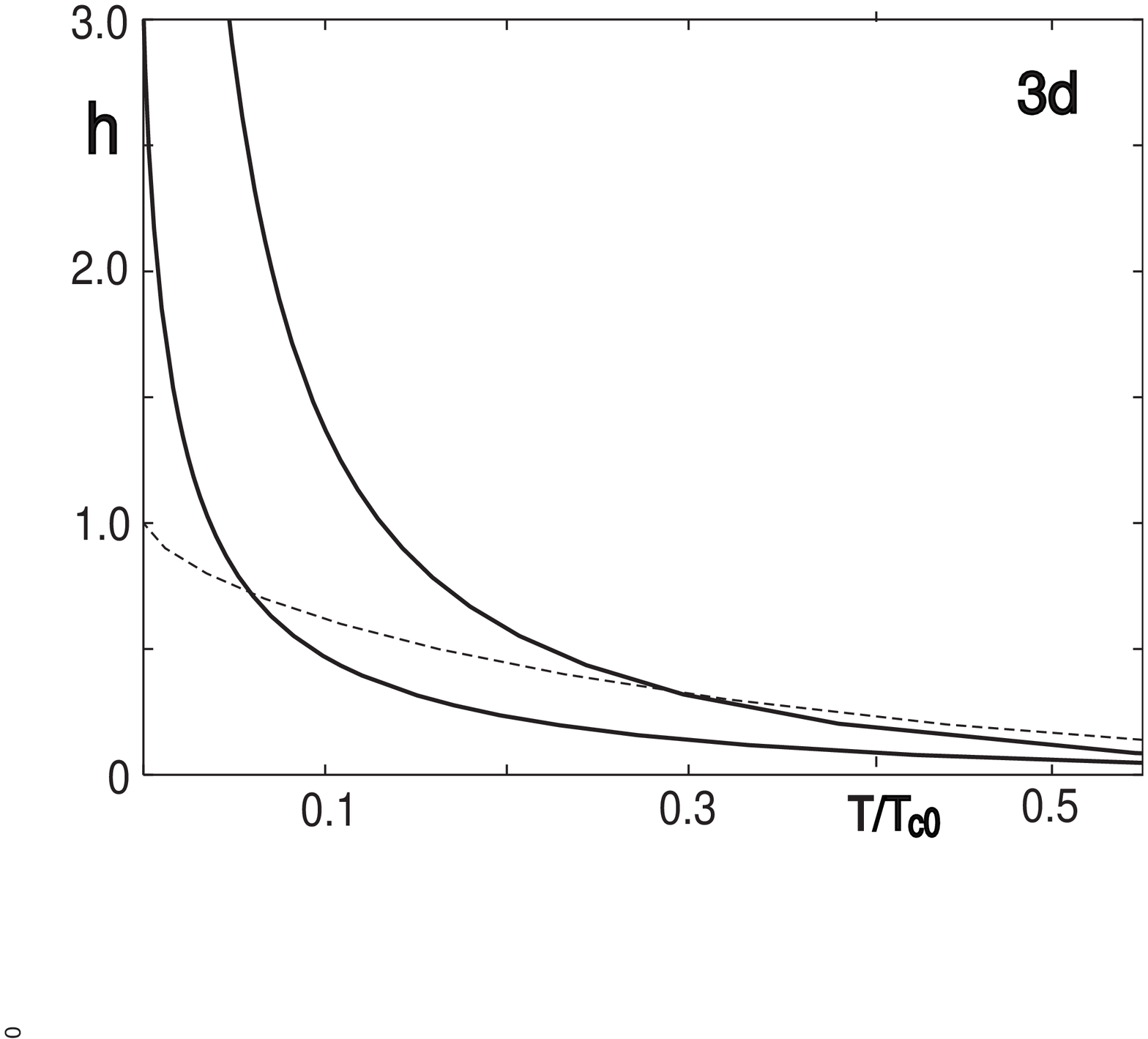}}
\caption{Examples of $H_{\rm vg}^{({\rm MF})}(T)$-lines in 3D case following from the present theory. The $H_{c2}(T)$ line is given by $h= 1 - T/T_{c0}$. See the text regarding the parameter values used for calculations. } \label{fig.3}
\end{figure}
%%%%%%%%%%%%%%%%%
%%%%%%%%%%%%%%%%%%%
\begin{figure}[t]
\scalebox{0.35}[0.35]{\includegraphics{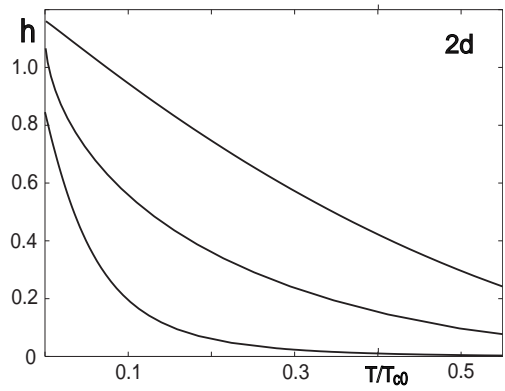}}
\caption{Examples of $H_{\rm vg}^{({\rm MF})}(T)$ in 2D (sharp crossover lines) comparable with those in Fig.1. See the text regarding the parameter values used for calculations. } \label{fig.4}
\end{figure}
%%%%%%%%%%%%%%%%%

Typical examples of computed $H_g(T)$ lines are shown in Fig.3 for 3D and Fig.4 for 2D, where $h=H/H_{c2}(0)$, and $T_{c0}$ is the zero field transition temperature. We have commonly used the parameter values, $2 \pi d_\psi J_0/t = 2 \pi \nu/32 = 10$ and $u_R/\alpha = 1 \times 10^{-3}$, and have changed $j_0=J_0/\alpha$ and $j=J/\alpha$. In Fig.3, the pairs of parameters ($j$, $j_0$) are ($0.05$, $0.7$) (left curve) and ($0.28$, $0.7$) (right), while in Fig.4 they were chosen as ($0.003$, $0.3$) (left curve), ($0.05$, $0.7$) (center), and ($0.05$, $0.3$) (right). The dashed curve in 3D is the thermal melting line in clean limit. 

In 3D case, the resistivity should vanish at $H_g(T)$, because, as is explained in sec.IV, the obtained $H_g(T)$ is essentially the same as the VG transition field \cite{FFH,RI2} in the amorphous case. In contrast, $H_g(T > 0)$ in 2D will be regarded as a crossover line along which the resistivity remarkably 
drops \cite{GaL}, while $H_g(0)$ may be identified with the SI transition field $H_{\rm SI}$ at $T=0$ (see sec.I). 

In 3D, the low $T$ limit $H_g(T \to 0)$ of the glass transition field lies, for most of parameter values we have examined, in $h > 1$, i.e., above the mean-field $H_{c2}(0)$. On the other hand, 
all values of the disorder strength $j$ used in the figures satisfy $|{\overline D}_{\omega=0}| > 0$. This implies assuming that, even at $T=0$, the PG order in $H > H_{c2}(0)$ does not occur. That is, Fig.3 implies that $H_g$ can lie above $H_{c2}(0)$ more easily than the prediction from the mean field analysis \cite{JL}. The presence of $H_g(T)$ above $H_{c2}(T)$ at low $T$ is due primarily to the coupling, appearing through ${\overline D}_{\omega=0}^{(d)}$, between the glass field and the SC fluctuation: For brevity, let us imagine that the SC fluctuation term in $({\overline D}_{\omega=0}^{(d)})^2$ (i.e., its second term) will be relatively small in magnitude. Then, from eq.(\ref{eq:38}) or (\ref{eq:39}), $\mu^{(d)}_0$ at $H_g(T)$ approximately scales like $(h^{(d)})^{2/(6-d)}$. Thus, the second term, $\sim h^{(d)}/(\mu^{(d)}_0)^{(4-d)/2}$, in $({\overline D}_{\omega=0}^{(d)})^2$ {\it grows} like $(h^{(d)})^{2/(6-d)}$ with increasing field $h^{(d)}$. As is explained in sec.V, this implies that the strength of vortex pinning is enhanced by the SC fluctuation with increasing field, although the relation $\mu^{(d)}_0 \sim (h^{(d)})^{2/(6-d)}$ itself implies a {\it reduction} of the SC fluctuation with increasing field above $H_{c2}$. Such an enhancement of pinning due to the SC fluctuation is peculiar to the granular superconductors at low $T$ and in high fields. In fact, at high $T$ and in weak or intermediate fields, $\mu^{(d)}_0$ and hence, $|{\overline D}_{\omega=0}^{(d)}|$ rather increases with increasing temperature. 

Further, it can be seen from the coupled equations leading to the figures that, in the present quantum model in which the main bare energy scale is not the Josephson coupling $J_0$ but the charging energy $\alpha$, the main $J_0$-dependence appears in $d_\psi \alpha/(J_0 t) \propto j_0^{-2}$, and hence that a larger $j_0$ leads to an {\it enhancement} of quantum fluctuation \cite{RIr,com6}. For this reason, the ($0.05$, $0.7$) curve lies below the ($0.05$, $0.3$) curve in Fig.4. 
\section{Review of results in nongranular case}

For the purpose of understanding the content of results in sec.III better, it is useful to compare the results in granular case with those in nongranular case \cite{RI2,Dorsey}. Here, we will sketch the corresponding analysis for obtaining the VG transition curve $H_g(T)$ of the amorphous-like materials. Within LLL, the familiar GL action derived microscopically takes the form 
\begin{eqnarray}\label{eq:Ap1}
{\cal S}_{GL}[\Psi] &=&  \int d^3r \int d\tau \biggl[ \biggl( ( \, \Psi({\bf r}, \tau) \, )^* \, u({\bf Q}; {\bf r}) \, \Psi({\bf r}, \tau)  \\ \nonumber 
&+& \xi_0^2 |\partial_z \Psi({\bf r}, \tau)|^2 
+ \frac{b}{2} |\Psi({\bf r}, \tau)|^4 \biggr) \\ \nonumber 
&+& \frac{\gamma}{2 \pi} \int d\tau' \frac{|\Psi({\bf r}, \tau) - \Psi({\bf r}, \tau')|^2}{(\tau - \tau')^2} \biggr], 
\end{eqnarray}
where ${\overline {u}} = {\rm ln}h$, $h = H/H_{c2}(0)$, 
${\overline {(u({\bf r}) - {\overline u}) (u({\bf r}') - {\overline u})}} = \Delta_0 \delta^{(3)}({\bf r} - {\bf r}')$, $b>0$, $\gamma > 0$, and ${\bf Q}$ is the gauge-invariant gradient in directions perpendicular to the applied field $\parallel {\hat z}$. The role of the random potential leading to the vortex pinning is played by $u({\bf r}) - {\overline u}$. The replicated action, arising after the random-average, takes the form
\begin{eqnarray}\label{eq:sglr}
{\cal S}_{\rm GL, r} &=& \sum_{a=1}^n {\cal S}_{\rm GL}[\Psi^{(a)}] 
\\ \nonumber 
&-& \frac{\Delta_0}{2} \int d^3r \sum_{a,b} \int d\tau \int d\tau' |\Psi^{(a)}({\bf r}, \tau)|^2 |\Psi^{(b)}({\bf r}, \tau')|^2. 
\end{eqnarray}
Alternatively, the last term of ${\cal S}_{\rm GL, r}$ may be regarded as arising from the action  
\begin{eqnarray}\label{eq:sglq}
{\cal S}_{{\rm GL},Q} &=& \int d\tau \int d\tau' \sum_{a,b} \int d^3r \biggl[ \frac{1}{2 \Delta_0} \delta Q^{(ab)}({\bf r}; \tau, \tau') \\ \nonumber 
&\times& \delta Q^{(ba)}({\bf r}; \tau', \tau) - (\Psi^{(a)}(\tau))^* \delta Q^{(ab)}(\tau, \tau') \Psi^{(b)}(\tau') \biggr]
\end{eqnarray}
after integrating over $\delta Q$. If restarting from the action $\sum_{a=1}^n {\cal S}_{\rm GL}[\Psi^{(a)}] + {\cal S}_{{\rm GL},Q}$, we can follow a similar route for obtaining the glass transition line to that in sec.III. First, 
the propagator of renormalized $\Psi$-fluctuation in LLL is specified by the Matsubara frequency $\omega$ and, in 3D case, the wave number $k$ in the direction of the applied field, and given in the form $( \mu + \gamma |\omega| + \xi_0^2 k^2)^{-1}$ in the Hartree approximation \cite{Dorsey}. For weak disorder, the parameter $\mu$ satisfies  
\begin{eqnarray}\label{eq:Ap2}
\mu &=& {\rm ln}h + b \, h \, \xi_0^{-3} \, T \sum_{\omega} \frac{1}{\sqrt{\mu + \gamma|\omega|}}, 
\end{eqnarray}
which corresponds to eq.(\ref{eq:44}) in sec.III. The VG transition field can be defined, just as in eq.(\ref{eq:37}), as a critical point of the glass fluctuation $\delta Q$ at which the inverse of the propagator $\langle \delta Q^{(ab)}({\bf k}; \omega_1, \omega_2) \delta Q^{(ba)}({\bf k}; -\omega_2, -\omega_1) \rangle$ vanishes in the limit of vanishing ${\bf k}$ and $\omega_j$. In the present case, the VG transition line is given by 
\begin{equation}\label{eq:Hvg}
1 = \frac{\Delta h}{2 \mu^{3/2}} 
\end{equation}
which corresponds to eq.(\ref{eq:38}) in the granular case, where $\Delta = \Delta_0/(2 \pi \xi_0^3)$. 

In previous works \cite{RI2,Dorsey}, eq.(\ref{eq:Hvg}) was obtained as a pole of the VG susceptibility directly constructed from the action (\ref{eq:sglr}). Then, the VG susceptibility takes the same form as the glass fluctuation propagator $\langle \delta Q^{(ab)}({\bf k}; \omega_1, \omega_2) \delta Q^{(ba)}({\bf k}; -\omega_2, -\omega_1) \rangle$ in the limit of vanishing ${\bf k}$ and $\omega_j$. That is, if starting the analysis in sec.III from an effective action composed only of $\psi$ corresponding to eq.(\ref{eq:sglr}), the glass transition and resistive behavior near $H_g(T)$ for the granular case can be described in the same manner as those performed elsewhere \cite{RI2,Dorsey}. Since the continuous vanishing of resistivity at the VG transition was explained there based on eq.(\ref{eq:sglr}), $H_g(T)$ for the 3D granular case in sec.III has to be also the SC transition line at which the resistivity continuously vanishes. 

The VG transition field in 3D case and at $T=0$ is determined by the following expression which is obtained from eqs.(\ref{eq:Ap2}) and (\ref{eq:Hvg}): 
\begin{eqnarray}\label{eq:Ap3}
{\rm ln} h &+& \frac{\pi^{-1} \xi_0^{-3} b \, \omega_c \, h}{ (h \Delta/2)^{1/3} + ( \gamma \omega_c + (h \Delta/2)^{2/3} )^{1/2} } \\ \nonumber 
&=& \biggl(\frac{h \Delta}{2} \biggr)^{2/3}, 
\end{eqnarray}
where $\omega_c$ is a high frequency cut-off. We note that the fluctuation-corrected (i.e, renormalized) value of $H_{c2}$, $H_{c2}^{(R)}$, is nonzero in 3D systems at $T=0$ and is given by eq.(\ref{eq:Ap3}) with $\Delta=0$. Due to this fact and the $\Delta$-dependence of eq.(\ref{eq:Ap3}), the 3D VG transition field at $T=0$ is always higher than $H_{c2}^{(R)}(0)$ and approaches $H_{c2}^{(R)}(0)$ as the disorder $\Delta$ diminishes \cite{IR,com9}. In contrast, in 2D case, $H_{c2}^{(R)}(0)$ vanishes, and hence, the VG transition field at $T=0$ (i.e., $H_{\rm SI}$) may lie below the mean field $H_{c2}(0)$. 

\section{Summary and Discussions}

First, let us start from explaining the 3D $H_g(T)$ going beyond the mean field $H_{c2}$ upon cooling in sec.III on the basis of the results in sec.IV for the nongranular case. One reason for the $H_g$-growth in higher fields seen in 3D case is that $H_g(0) > H_{c2}^{(R)}(0)$ in any 3D case, and that $H_{c2}^{(R)}(0)$ is nonvanishing and tends to lie near the mean field $H_{c2}(0)$. For instance, even in the left curve in Fig.3 where $T_{g}(H)$ is significantly lowered, $H_{c2}^{(R)}(0)$ lies below but close to $H_{c2}(0)$. For completeness, we show the lowest $T$ behaviors of $H_g(T)$ line for extremely low $j$ values in Fig.5. It shows that, as mentioned below eq.(\ref{eq:Ap3}) in the nongranular case, $H_g(0)$ approaches $H_{c2}^{(R)}(0) = 0.978 \, H_{c2}(0)$ as the disorder $j$ diminishes. That is, since $H_{c2}^{(R)}(0)$ is the lower limit of the 3D SC (i.e., glass) transition field at $T=0$, $H_{g}(T \to 0)$ in $d=3$ lies in $h > 1$ in most cases (see also Appendix). In contrast, $H_{c2}^{(R)}(0)$ in 2D case is zero, and consequently, $H_{\rm SI}$ of a system with strong enough quantum fluctuation, as the left curve in Fig.3 shows, can lie in $h < 1$, i.e., much below the mean field $H_{c2}(0)$. This situation corresponds to the case in which a field-tuned S-I transition behavior is seen through resistivity curves in granular materials \cite{Gant}. 
%%%%%%%%%%%%%%%%%%%
\begin{figure}[t]
\scalebox{0.35}[0.35]{\includegraphics{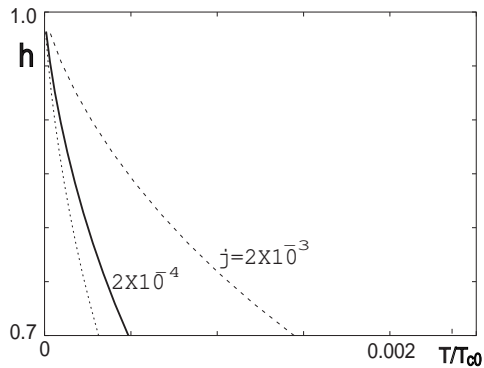}}
\caption{3D $H_g(0)$ curves at very low $t=T/T_{c0}$ for the weak disorder cases with $j = 1 \times 10^{-4}$, $2 \times 10^{-4}$, and $2 \times 10^{-3}$. In common, the value $j_0=1.5$ was used. } \label{fig.5}
\end{figure}
%%%%%%%%%%%%%%%%%

More importantly, in the granular case at low $T$, there is a contribution of the SC fluctuation {\it enhancing} the vortex pinning strength with increasing field: As explained in sec.III, the l.h.s. of each of eqs.(\ref{eq:38}) and (\ref{eq:39}) decreases with increasing field due to the SC fluctuation. On the other hand, by comparing eq.(\ref{eq:38}) with eq.(\ref{eq:Hvg}), the inverse of the l.h.s. of eq.(\ref{eq:38}), increasing with field, corresponds to a strength of vortex pinning inducing the glass transition. This unfamiliar fluctuation effect, peculiar to granular systems, is quantitatively weakened in 2D case, because the quantum fluctuation is stronger in lower dimensions and selfconsistently increases $\mu^{(d)}_0$. 

In the present paper, we have examined the low $T$ behavior of SC glass transition curve $H_g(T)$ of granular superconductors by applying a theory of quantum spin-glass to the context of superconductivity and have shown that, in contrast to the situations in amorphous-like materials, a situation with $H_g > H_{c2}$ at low $T$ usually occurs in 3D granular systems. This is consistent with the phenomenological picture \cite{GaL,Gesh,JL} favoring the presence of a {\it superconducting} glass in $H > H_{c2}(0)$. Next, we have shown that, without the PG order, the {\it fluctuation} of the glass order parameter plays the role of pinning disorder inducing a vortex-glass instability at $T > 0$ in 3D systems under nonzero fields. 

Our main messages based on the results in sec.III and IV is that the SC transition in nonzero fields can be described in a single theory, i.e., as a vortex-glass transition \cite{FFH}, for both the granular and nongranular (amorphous-like) materials. Because the granular and amorphous-like systems may be continuously connected with each other, e.g., by changing a composition of materials, such a unified view of two limiting models of disordered superconductors should be naturally expected. In a work with a similar purpose to the present one, Galitski and Larkin \cite{GaL} have argued that even the SC transition in amorphous-like materials should be described within a model for granular systems. Our result in sec.III that $H_{c2}(0) < H_g(0) < +\infty$ is consistent with the argument in Refs.\cite{GaL} and \cite{Gesh}. However, our result in 2D case, given in Fig.4, that $H_{g}(T \to 0) < H_{c2}(0)$ is different from their opinion and rather consistent with experimental facts showing the S-I transition behavior. 

The vanishing of resistivity on approaching $H_g$ from above should imply that the glass phase in $H < H_g$ is superconducting, because the transition at $H_g$ is continuous in the present case. It is not surprising that the present result disagrees with an argument \cite{baka} based on almost the same model that even the 3D VG is a metal, because the dissipative term of eq.(\ref{eq:2}) was not taken into account correctly in Ref.\cite{baka}. Without the dissipative term, a finite fluctuation conductivity at finite $T$ above $H_g$ cannot be discussed. \cite{RIr} In fact, it is difficult to reconcile the metallic response \cite{baka} in PG phase with the conventional ac Meissner response \cite{Feigel}. Further, if their model with no Ohmic dissipation \cite{baka} is extended to the case with correlated line-like disorder creating the so-called Bose glass phase \cite{Bla}, a (Bose) glass phase with {\it finite} resistivity seems to be still obtained in contrast to experimental facts. Our analysis on the conductivity in glass phases including these results will be presented separately. \cite{RI4} 

\begin{acknowledgements}
This work is finantially supported by a Grant-in-Aid from the Ministry of Education, Culture, Sports, Science, and Technology, Japan. 
\end{acknowledgements}

\vspace{5mm}

\section{Appendix}

Here, we estimate $H_g(0)$ in 3D amorphous-like case at low $T$ limit by using microscopic parameters in dirty limit. 
In the ordinary dirty limit where $T \tau \ll 1$, the coefficients in the action (\ref{eq:Ap1}) are available \cite{RIr,RI3,Vojta}. If reasonably choosing $\gamma \omega_c$ to be a constant of order unity, we have 
$\omega_c \sim \tau^{-1}$, where $\tau$ is the elastic scattering time of quasiparticles, and 
\begin{eqnarray}
b \frac{\omega_c}{\xi_0^3} &\simeq& \biggl(\frac{\xi_0^{({\rm cl})}}{\xi_0} \biggr)^3 \biggl( \frac{T_c}{E_{\rm F}} \biggr)^2 \frac{1}{T_c \tau}, \\ \nonumber
\Delta &\simeq& b \frac{\omega_c}{\xi_0^3} \frac{1}{E_{\rm F} \tau}, 
\end{eqnarray} 
where $\xi_0^{({\rm cl})}$ is the $T=0$ coherence length in clean limit, $T_c$ is the zero field transition temperature, and $h$ dependences in the coefficients were neglected by assuming the glass transition field $H_g(0)$ in $T \to 0$ limit to stay close to $H_{c2}(0)$. This is justified as far as both $b \omega_c/\xi_0^3$ and $\Delta \ll 1$. If $T_c \tau \ll 1$, $b \omega_c/\xi_0^3 \gg \Delta^{2/3}$, and consequently, $H_g(0)$ lies below $H_{c2}(0)$. By contrast, for larger $\tau T_c$ values of order unity, $H_g(0)$ might lie rather above $H_{c2}(0)$. Nevertheless, its difference $(H_g(0) - H_{c2}(0))/H_{c2}(0)$ is small according to 
\begin{equation}\label{eq:Aplast}
H_g(0) - H_{c2}(0) \sim H_{c2}(0) \biggl(\frac{T_c}{E_{\rm F}} \biggr)^2 \frac{1}{(T_c \tau)^{10/3}}. 
\end{equation}
Of course, when $T_c \tau \gg 1$, the rhs in eq.(\ref{eq:Ap3}) is negligible, and $H_g(0)$ becomes \cite{IR} of the order of the nonvanishing $H_{c2}^{(R)}(0)$ just below the mean field $H_{c2}(0)$. 

In the above analysis, effects of Coulomb repulsion \cite{com3} between quasiparticles were neglected. As in 2D case \cite{IR,RI1,RI3}, it would play a role of reducing $H_g(0)$, although $H_g(0)$ does not become lower than $H_{c2}^{(R)}(0)$ in 3D case. 

In this manner, the statement in Introduction that $H_g(0)$ of nongranular superconductors lies close to $H_{c2}(0)$ is justified. 

%\section{References}

%\begin{eqnarray}
%\mu_0 = \frac{\alpha}{J_0} \frac{r_{\psi, 0}}{t} + \frac{2 \sqrt{3} \pi h_3}{36%} + \Sigma - \alpha^{-1} {\overline D}_{\omega=0}
%\end{eqnarray}

%\begin{eqnarray}
%\mu_0 = \frac{\alpha}{J_0} \frac{r_{\psi, 0}}{t} + \frac{2 \pi h_2}{8} + \Sigma% - \alpha^{-1} {\overline D}_\omega
%\end{eqnarray}


\begin{thebibliography}{99}
\bibitem{com6} Alternatively, the strength of quantum fluctuation may be tuned by directly reducing $\nu$ with other parameters including $j_0$ fixed. Of course, $T_{c0}$ and $H_{c2}(0)$ increase with increasing $J_0$.
\bibitem{com9} Since we focus here on a narrow field range close to $H_{c2}(0)$ in clean 3D systems at $T=0$, effects of the solidification of the vortices occurring in clean systems (see sec.4 of Ref.6) will not be considered here.
\end{thebibliography}
\end{document}